Short communication

# Calotropin from milk of *Calotropis gigantean* a potent inhibitor of COVID 19 corona virus infection by Molecular docking studies

Arun Dev Sharma* and Inderjeet Kaur

*PG dept of Biotechnology, Lyallpur Khalsa College Jalandhar*

*Corresponding author, e mail: arundevsharma47@gmail.com*

ABSTRACT

SARS-CoV-2 (COVID-19), a positive single stranded RNA virus, member of corona virus family, is spreading its tentacles across the world due to lack of drugs at present. Being associated with cough, fever, and respiratory distress, this disease caused more than 15 % mortality worldwide. Due to its vital role in virus replication, Mpro/3CLpro has recently been regarded as a suitable target for drug design. The current study focused on the inhibitory activity of Calotropin, a component from milk of *Calotropis gigantean*, against Mpro protein from SARS-CoV-2. Till date there is no work is undertaken on in-silico analysis of this compound against $M^{pro}$ of COVID-19 protein. In the present study, molecular docking studies were conducted by using Patchdock tool. Protein Interactions tool was used for protein interactions. The calculated parameters such as docking score indicated effective binding of Calotropin to Mpro protein. Interactions results indicated that, $M^{pro}$/ Calotropin complexes forms hydrophobic interactions. Therefore, Calotropin may represent potential herbal treatment to act as COVID-19 $M^{pro}$ inhibitor. However, further research is necessary to investigate their potential medicinal use.

INTRODUCTION

COVID-19 is easily transmissible and it has already been spread worldwide. Symptoms are flu-like and can include fever, muscle and body aches, coughing, and sore throat. Symptoms may appear 5-6 days after infection. As of March 20h, 2020, over 243,000 cases of COVID-19 have been confirmed worldwide, over 10,000 of which have resulted in death. At present, no specific therapies for COVID-19 are available and research regarding the treatment of COVID-19 is infancy[1]. Some preliminary studies have investigated potential combinations that include anti malarial drug hydroxychloroquine, and anti-HIV vaccines can be used to treat COVID-19 infections. A separate investigation indicated that among 4 tested drugs Ritonavir was identified as the best potential inhibitor against COVID-19[2]. $M^{pro}$ also named as $3CL^{pro}$, from SARS-CoV-2, represents a potential target for the inhibition of CoV replication [1]. $M^{pro/}$3CL is involved in spike, membrane, envelop, nucleoprotein, replicase, and polymerase activity of viruses. Therefore, by virtue of its key role in polyprotein processing and virus maturation, Mpro is considered to be a suitable target for viral inhibitor development. In drug discovery processes, in silico drug designing is a form of computer-based modeling which is very useful. This method offers advantage to deliver a new drug in fast and cost-effective manner. As of this date, the process of drug designing has been advanced with various bioinformatics tools which helps in molecular docking analysis, protein- protein/ligand interaction, virtual screening and de-novo synthesis and in silico ADMET prediction [2]. From ancient time, medicinal plants are beneficial in the field of drug therapeutics as they are safer alternatives being utilized by humans for centuries. Previously, many of the new drug formulations are derived from natural products. Our present study focuses on the in silico analysis of Calotropin from milk of *Calotropis gigantean.* Calotropis, a member of Apocynaceae family, is a poisonous plant [3]. The active principles are uscharin, calotoxin, calactin, and calotropin. The leaves and stem when incised yield thick milky juice. Calotropis gigantea, the crown flower, is a species of Calotropis native to Cambodia, Bangladesh, Indonesia, Malaysia, the Philippines, Thailand, Sri Lanka, India, China, Pakistan, Nepal, and tropical Africa. It is a large shrub growing to 4 m (13 ft) tall. Milk of this plant has been used in various system of medicine for the past 2000 years. Bioactive components from leaves, flowers, fruits and roots of *Calotropis gigantean* are used as pesticide, insecticide, fungicide and Nematicidal/Schistosomicidal/Antihelminthic Activity, but little is known about its antiviral potential. We hypothesize that Calotropin from *Calotropis gigantean* has the capability to prevent infection of COVID-19. However, in the present study, we investigated Calotropin as potential inhibitor candidates for COVID-19 M$^{pro}$. Therefore the research objective of the present study was *in-silico* analysis and comparative molecular docking studies pertain to Calotropin in relation with M$^{pro}$ protein. The findings of the present study will provide other researchers with opportunities to identify the right drug to combat COVID-19.

MATERIAL AND METHODS

Proteins/Macromolecules

COVID-19 3CLpro/Mpro structures were obtained from PDB (https://www.rcsb.org/). The native ligand for 3clpro/Mpro structures was Calotropin

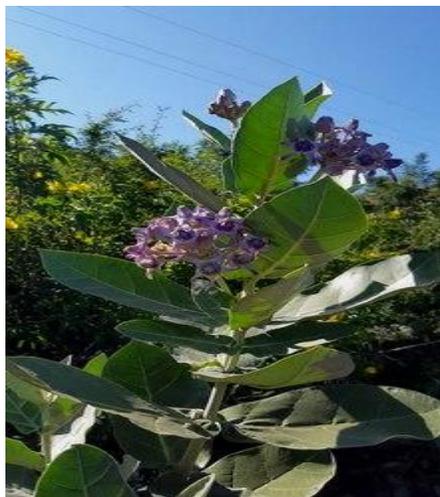

Fig 1: Pictorial image of Calotropis gigantea

Ligand and Drug Scan

The 3-dimensional (3D) structure of Calotropin, were obtained from PubChem (https://pubchem.ncbi.nlm.nih.gov/). PubChem is a chemical substance and biological activities repository consisting of three databases, including substance, compound, and bioassay databases.

Binding Mode of Docked Complexes

The docked complex structure output format was submitted into the Protein Interactions Calculator (PIC) webserver (https://projects.biotec.tu-dresden.de/plip-web/plip/index/) in order to map the interaction of the resulting docked complex. The parameters such as number of hydrogen bonds, number of hydrophobic residues, and number of aromatic and ionic interactions were considered in interpreting the strength of the interaction

Determination of Active Sites

The amino acids in the active site of a protein were determined using the Computed Atlas for Surface Topography of Proteins (CASTp) (http://sts.bioe.uic.edu/castp/index.html?201l).

Molecular Docking

PatchDock tool was used for docking study of the compounds over COVID-19 S-protein (https://bioinfo3d.cs.tau.ac.il/PatchDock/).

Admet SAR

AdmetSAR is a comprehensive web-based tool for predicting ADMET properties of candidate molecules (http://lmmd.ecust.edu.cn/admetsar2/). It predicts the pharmacokinetic properties such as Absorption, Distribution, Metabolism, Excretion and Toxicity.

RESULT AND DISCUSSION

SARS-CoV-2, member of corona viruses (CoVs) family, is enveloped, positive single stranded RNA viruse. $M^{pro}$) found in the SARS-Co-V-2 can be accessed in PDB and was suggested to be a potential drug target for 2019-nCov [1]. In many viruses, proteases play essential roles in viral replication; therefore, proteases are often used as protein targets during the development of antiviral therapeutics. In SARS-CoV-2, the $M^{pro}$ protein is involved in virus proteolytic maturation and has been examined as a potential target protein by inhibiting the cleavage of the viral polyprotein to prevent the spread of infection. The invention of the Mpro/3CLpro protease structure in COVID-19 provides a nice path to identify potential drug candidates to prevent infection. As cited earlier [2], proteases represent key targets for the inhibition virus replication, and the protein sequences of the SARS-CoV-2 $M^{pro}$ and the 2019-nCoV Mpro are 96% identical, hence host proteases can be used as potential therapeutic targets.

Traditionally, the discovery of new therapeutic drugs is a tedious and expensive process which generally takes 12-14 years with a lot of money to bring drug from discovery to market. In order to overcome these problems a lot many multidisciplinary approaches are used to discover new drug. In drug discovery processes, in silico drug designing is a form of computer-based modeling which is very useful. In the field of drug discovery, medicinal plants are advantageous as they are utilized as a safe herbal alternative by humans for centuries. The sources of many of the new drugs and active ingredients of medicines are derived from natural products. In the present study, we performed in silico analysis of Calotropin from *Calotropis gigantean* against $M^{pro}$ protein of SARS-CoV-2. In modern drug discovery process, molecular docking is a widely used computational method to predict the binding mode and binding affinity of ligands with the target receptor protein. The efficacy of the docked complex was evaluated on the basis of two essential criteria's: The minimum binding energy and the interaction of the ligand with the active site residues. A ligand undergoes either hydrogen bonding or hydrophobic interactions or both while docking into the active site. The results of docking can be used to find the best inhibitors for specific target proteins and thus to design new drugs. We followed the structural biology aspects which focus on the availability and retrieval of a main protease ($M^{pro}$) or 3C-like protease ($3CL^{pro}$) receptor structure from PDB database. The ligand Calotropin from *Calotropis gigantean* was docked to main protease (Mpro) or 3C-like protease (3CLpro). Previously several ligands and drug candidate compounds have been selected, as per criteria of Lipinski's rule of five. So the Calotropin that did not incur more than 2 violations of Lipinski's rule could be used in molecular docking experiments with the target protein. The drug scanning results (data not shown) showed that Calotropin from *Calotropis gigantean*, herbal ligand used in this study, was accepted by Lipinski's rule of five.

Molecular docking using 1-click docking tool that was used to find out interaction of inhibitor i.e Calotropin from *Calotropis gigantean* with $M^{pro}/3CL^{pro}$ protein revealed 10 different poses

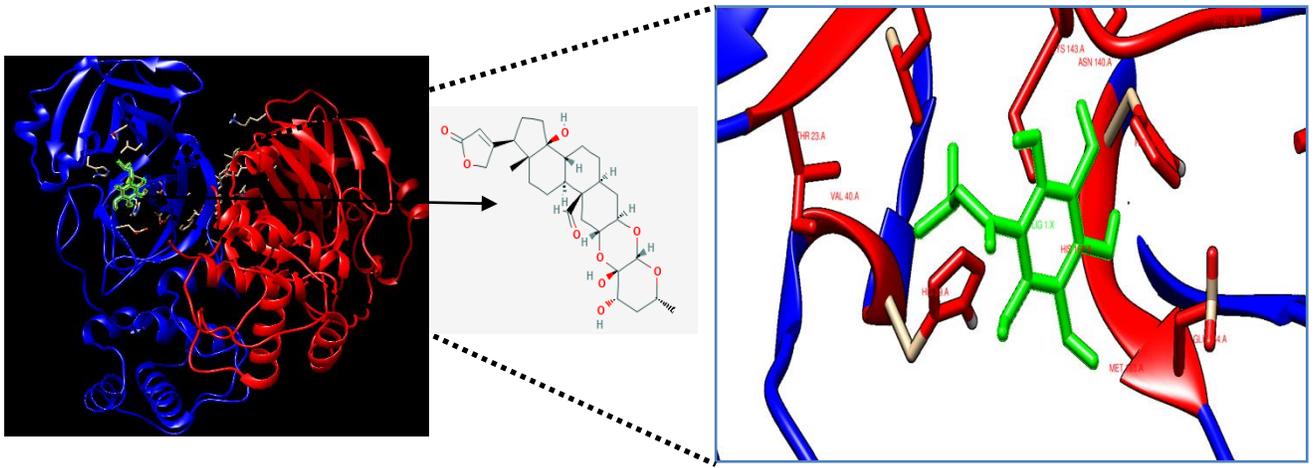

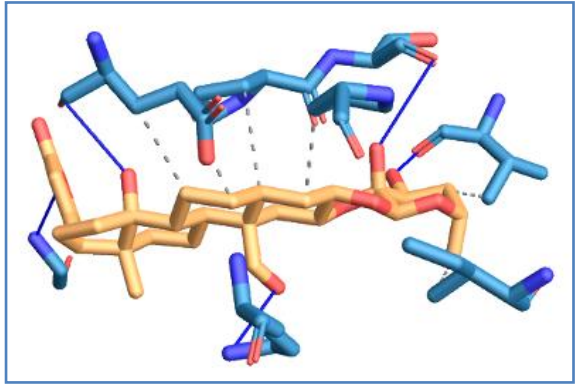

Fig 1: Molecular docking pose and interactions of ligand with Mpro protein and 2D interactions of Calotropin

Table 1. Binding Energy and full fitness values of Docked complex and interaction values

| Inhibitor | Dock pose | Dock score | Active site residues | |
|---|---|---|---|---|
| | | | H-bond interactions | Hydrophobic interactions |
| Calotropin | 1 | -253.66 | LYS10, GLU14, GLY71 | PRO7, 122, GLU114, ILE150, VAL301 |

based on the dock score and the pose with highest negative values indicated maximum binding affinity. Best docking pose and molecular interactions of Calotropin in Fig. 1. Calotropin was successfully docked with $M^{pro}$ protein at binding pocket S1 with dock score -253.66 (Table 1). The interaction of Calotropin in the binding pocket of $M^{pro}$ protein was mediated by two hydrophobic interactions and hydrogen bond interactions (Table 1). The Calotropin showed full fitness within active site amino acids of Mpro/3CLpro proteins of COVID-19 (Fig. 1). We investigated that Calotropin as potential inhibitor of the SARS-CoV-2 $M^{pro}$. Hydroxy groups (-OH), ketone groups (=O) and ether groups (-O-) in Calotropin compounds are predicted to play roles amino acid residue interactions at the active site of SARS-CoV-2 $M^{pro}$. Finally, lack of wet-lab validation is a drawback in our research and we expect computational biology analysis and its integration with wet-lab data can be productive in the determination of potential anti-Mpro/3CLpro components. In silico pharmacokinetic analysis of eucalyptus was conducted using Admet SAR. The pharmacokinetics of a drug depends on its absorption, distribution, metabolism, excretion and toxicity. Calotropin molecule showed positive results for for BBB profiles. Calotropin compound was non-substrate to Pgp (data not shown)

CONCLUSION

Due to non approved drugs at present Currently, SARS-CoV-2 has emerged in the human population, in China, and is a potential threat to global health, worldwide. Currently, the main target for SARS-CoV-2 treatment primarily act on the main protease (Mpro). The aim of this study was to examine Calotropin that may be used to inhibit the SARS-CoV-2 infection pathway. Therefore, we suggested that Calotropin may represent potential herbal treatment options, and found in medicinal plants that may act as potential inhibitors of SARS-CoV-2 M$^{pro}$. However, further studies should be conducted for the validation of these compounds using in vitro and in vivo models to pave a way for these compounds in drug discovery.

ACKNOWLEDGMENT

ADS want to thank management for this support.

CONFLICT OF INTEREST

Authors declares no conflict of interest

COMPLIANCE WITH ETHICAL STANDARDS

The authors declare that they have no conflict of interest. This article does not contain any studies involving animals or human participants performed by any of the authors

AUTHOR CONTRIBUTIONS

ADS: designed the study and prepared manuscript

IJK: designed the study and prepared manuscript

REFERENCES

1. Lu H. Drug treatment options for the 2019-new coronavirus (2019-nCoV), *Biosci. Trends*, (2020), doi:10.5582/bst.01020.
2. Liu X. and Wang X.J., Potential inhibitors against 2019-nCoV coronavirus M protease from clinically 13 of 14 approved medicine, *J. Genet. Genomics*, (2020), doi: 10.1016/j.jgg.2020.02.001.
3. Wang, Shih-Chung; Lu, Mei-Chin; Chen, Hsiu-Lin; Tseng, Hsing-I; Ke, Yu-Yuan; Wu, Yang-Chang; Yang, Pei-Yu (2009). "Cytotoxicity of calotropin is through caspase activation and down regulation of anti-apoptotic proteins in K562 cells". Cell Biology International. 33 (12): 1230–1236.

4. Im K., Kim J. and Min H. Ginseng, the natural effectual antiviral: Protective effects of Koran Red Ginseng against viral infection, Retrieved from https://www.ncbi.nlm.nih.gov/pmc/articles/PMC5052424/, (2015).
5. Rodríguez-Morales P., Alfonso J., MacGregor K., Kanagarajah S. and Dipti P. Going global – Travel and the 2019 novel coronavirus, *Travel Med. Infect. Dis*., 33, (2020), doi: https://doi.org/10.1016/j.tmaid.2020.101578.